\documentclass[aps,showpacs,preprintnumbers,nofootinbib,preprint]{revtex4}

\usepackage{amsmath}
\usepackage{amssymb}
\usepackage{graphicx}

\begin{document}

\title{Dark energy as a massive vector field}

\author{C. G. B\"ohmer}
\email{christian.boehmer@port.ac.uk}
\affiliation{Institute of Cosmology and Gravitation,
             University of Portsmouth, Portsmouth, PO1 2EG, UK}

\author{T. Harko}
\email{harko@hkucc.hku.hk} \affiliation{Department of Physics and
Center for Theoretical and Computational Physics, The University
of Hong Kong, Pok Fu Lam Road, Hong Kong SAR, P. R. China}
\date{January 11, 2007}

\begin{abstract}
We propose that the Universe is filled with a massive vector
field, non-minimally coupled to gravitation. The field equations
of the model are consistently derived and their application to
cosmology is considered. The Friedmann equations acquire an extra
dark-energy component, which is proportional to the mass of the
vector particle. This leads to a late-time accelerated de Sitter type
expansion. The free parameters of the model (gravitational
coupling constants and initial value of the cosmological vector
field) can be estimated by using the PPN solar system constraints.
The mass of the cosmological massive vector particle, which may
represent the main component of the Universe, is of the order of
$10^{-63}$ g.
\end{abstract}

\pacs{03.70.+k, 11.90.+t, 11.10.Kk}

\maketitle

\section{Introduction}

An increasing number of observational data, obtained in the past few years,
strongly support a model according to which the Universe is spatially flat,
mostly made of non-conventional matter -- baryons being allowed only up to $4\% $
of the total energy content -- and accelerating. The physical
models accounting for such a picture generally contain two basic
ingredients: pressureless dark matter (DM), responsible for the
growth of cosmological perturbations via gravitational
instability, and negative pressure dark energy (DE), responsible
for the accelerated expansion (for recent reviews on the dark
energy problem see~\cite{rev}).

The simplest model along these lines is $\Lambda $CDM, in which the role of
DE is played by a cosmological constant $\Lambda $. It fits very well all
the data related with the cosmological background and the perturbations in
the linear regime~\cite{sel05}. The cosmological constant is attributed to
the quantum zero-point energy of the particle physics vacuum, with a
constant energy density $\rho $, pressure $p$ and an equation of state $%
w=p/\rho =-1$. Existing observational data indicate the equation of state of
dark energy very close to the cosmological constant value, $w=-1\pm 0.2$ at $%
95\%$ confidence level, with at most a very mild evolution up to redshift $%
z\sim 1$.

An alternative model to the cosmological constant may be the
quintessence~\cite{quint}, a dynamical scalar field which, in the
simplest model, slowly rolls in a potential characterized by an
extremely low mass. The spatially averaged equation of state for
the quintessence field satisfies $w>-1$. In order to avoid
fine-tuning on the initial conditions, the quintessence scalar
field is usually taken to be extremely light, with a Compton
wavelength corresponding to the present value of the Hubble
radius. As a consequence, the scalar field is homogeneous on all
observable scales, much like a cosmological constant. However,
quintessence models suffer from an important theoretical problem,
namely, the fact that radiative corrections induced by the
couplings with the matter fields would generically induce huge
corrections to the trace level mass, thus spoiling the required
lightness. Hence to keep the scalar field light in these models a
fine-tuning on the radiative corrections is generally required,
besides the one necessary to keep the cosmological constant small.

A wide variety of other dark energy models has also been proposed,
including K-essence~\cite{K}, Chaplygin gas~\cite{chap},
modifications of gravity~\cite{grav}, Born-Infeld scalars (rolling
tachyon)~\cite{born}, massive scalars~\cite{scal} etc. The common
feature of these models is that they operate through an
undetermined field potential which in principle can incorporate
any a priori associated cosmological evolution, thus lacking
predictive power at the fundamental level~\cite{born}. There is a
tremendous degeneracy in these models and generally they are
judged by their physical implications and by the generic features
which arise in them. Therefore a consistent physical picture of
the dark energy, which could explain the size of its energy
density $\rho _{{\rm DE}}\approx 10^{-12}\;{\rm eV}^{4}$, and
suggests how the underlying physics may be probed, is still
missing.

Non-gravitational interactions are known to be mediated by vector
fields. Therefore the possibility that a vector field, which, for
example, may be a partner of quintessence, could be at the origin
of the present stage of the cosmic acceleration cannot be
neglected. Theoretical proposals in which a minimally coupled
vector field is responsible for the present dynamics of the
Universe have been considered in~\cite{mod}.

It is the purpose of the present paper to propose a model of the
dark energy in terms of a massive, Proca type vector field, with a
non-minimal coupling to the gravitational field. The model
contains three independent
parameters $\omega $, $\eta $ and $\mu _{\Lambda }^{2}$, respectively, with $%
\mu _{\Lambda }^{2}$ representing the mass of the massive
cosmological vector particle. The scalars $\omega $ and $\eta $
describe the non-minimal coupling of the vector field to the Ricci
scalar and to the Ricci tensor, respectively. The gravitational
field equations can be consistently derived from a variational
principle. A similar vector-tensor theory, without the mass term,
was also proposed in the early 1970's~\cite{hel73}.

In the cosmological case, corresponding to a flat homogeneous and
isotropic Universe, the Friedmann equations acquire an extra
dark-energy component, which is proportional to the mass of the
vector particle. This term, playing the role of the cosmological
constant, leads to the late accelerated expansion of the Universe.
The free parameters of the model (gravitational coupling constants
and initial value of the cosmological vector field) can be
estimated by using the PPN solar system constraints. The mass of
the cosmological massive vector particle, which may represent the
main component of the Universe, is of the order of $10^{-63}$ g.

The present paper is organized as follows. The field equations of
the massive vector-tensor theory are derived in Section II. In
Section III we consider the cosmological applications of the
model. The PPN constraints on the model parameters are discussed
in Section IV. We discuss and conclude our results in Section V.

Throughout this paper we use the Landau-Lifshitz conventions
\cite{LaLi} for the metric signature $\left( +,-,-,-\right) $ and
for the field equations, and a system of units with $c=\hbar =1$.

\section{Field equations of the massive vector-tensor theory}

We assume that the Universe is filled with a massive cosmological
vector field, with mass $\mu _{\Lambda }$, which is characterized
by a four-potential $\Lambda ^{\mu }\left( x^{\nu }\right) $, 
$\mu,\nu =0,1,2,3$ and which couples non-minimally to gravity. In
analogy with electrodynamics we introduce the field tensor
\begin{equation}
C_{\mu \nu }=\nabla _{\mu } \Lambda _{\nu }-\nabla _{\nu } \Lambda_{\mu }.
\end{equation}

The interaction of the gravitational and of the vector fields is
described by a Lagrangian which is required to satisfy the
following conditions: a) the Lagrangian density is a four-scalar
b) the free-field energies are positive-definite for both the
metric and the vector field c) the resulting theory is metric and
d) the field equations contain no higher than second derivatives
of the fields~\cite{hel73}. The action for such a theory can be
written as
\begin{equation}
S=-\int \left[ R+C_{\mu \nu }C^{\mu \nu }+\frac{1}{2}\mu _{\Lambda
}^{2}\Lambda _{\mu }\Lambda ^{\mu }+\omega \Lambda _{\mu }\Lambda ^{\mu
}R+\eta \Lambda ^{\mu }\Lambda ^{\nu }R_{\mu \nu }+16\pi G_0L_{m}\right] \sqrt{%
-g}d\Omega ,  \label{1}
\end{equation}
where $R_{\mu \nu }$ and $R$ are the Ricci tensor and the Ricci
scalar, respectively, $G_0$ is the gravitational constant and
$L_{m}$ is the matter Lagrangian. In Eq.~(\ref{1}) $\omega $ and
$\eta $ are dimensionless coupling parameters. The
four-dimensional volume element is $d\Omega
=dx^{0}dx^{1}dx^{2}dx^{3}$. In the following we denote $\phi
=\Lambda _{\mu }\Lambda ^{\mu }$, which is an invariant scalar.

The variation of the action with respect to the metric tensor $g_{\mu \nu }$
gives the field equations
\begin{eqnarray}\label{field}
R_{\mu \nu }-\frac{1}{2}g_{\mu \nu }R+\omega \left[ \phi \left(
R_{\mu \nu }-\frac{1}{2}g_{\mu \nu }R\right) +\Lambda _{\mu
}\Lambda _{\nu }R+g_{\mu \nu }\nabla _{\lambda }\nabla ^{\lambda
}\phi -\nabla _{\nu }\nabla _{\mu }\phi
\right] +\nonumber\\
 \eta \Lambda ^{\alpha }\Lambda ^{\beta
}\left( g_{\mu \alpha }R_{\nu \beta }+g_{\nu \beta }R_{\mu \alpha
}-\frac{1}{2}g_{\mu \nu }R_{\alpha \beta }\right) +  \nonumber\\
\frac{\eta }{2}\left[ g_{\mu \nu }\nabla _{\alpha }\nabla _{\beta
}\left( \Lambda ^{\alpha }\Lambda ^{\beta }\right) +\nabla
_{\sigma }\nabla ^{\sigma }\left( \Lambda _{\mu }\Lambda _{\nu
}\right) -\nabla _{\sigma }\nabla _{\nu }\left( \Lambda _{\mu
}\Lambda ^{\sigma }\right) -\nabla _{\sigma }\nabla _{\mu }\left(
\Lambda _{\nu
}\Lambda ^{\sigma }\right) \right] +\nonumber\\
2C_{\mu \sigma }C_{\nu }^{\sigma }-\frac{1}{2}g_{\mu \nu
}C_{\alpha \beta }C^{\alpha \beta }+\frac{1}{2}\mu _{\Lambda
}^{2}\Lambda _{\mu }\Lambda _{\nu }-\frac{1}{4}\mu _{\Lambda
}^{2}\phi g_{\mu \nu }=8\pi G_0T_{\mu \nu },
\end{eqnarray}
where $T_{\mu \nu }$ is the energy-momentum tensor of the matter, defined in
terms of the matter action $S_{m}=16\pi G_0\int L_{m}\sqrt{-g}d\Omega $ as $%
\delta S_{m}=8\pi G_0\int T_{\mu \nu }\delta g^{\mu \nu
}\sqrt{-g}d\Omega $.

The variation of the action with respect to $\Lambda _{\mu }$ gives the
generalized Maxwell equation
\begin{equation}
4\nabla _{\nu }C^{\mu \nu }+\omega R\Lambda ^{\mu }+\eta \Lambda
^{\nu }R_{\nu }^{\mu }+\frac{1}{2}\mu _{\Lambda }^{2}\Lambda ^{\mu
}=0. \label{3}
\end{equation}

By contracting the field equations we find
\begin{equation}
-R+\left( 3\omega +\frac{\eta }{2}\right) \nabla _{\lambda }\nabla
^{\lambda }\phi +\eta \nabla _{\alpha }\nabla _{\beta }\left(
\Lambda ^{\alpha }\Lambda ^{\beta }\right)=8\pi GT,
\end{equation}
where $T=T_{\mu }^{\mu }$. Taking the covariant derivative of Eq.~(\ref{3})
we obtain the conservation law for the four-potential of the massive
cosmological field as
\begin{equation}
\left( \frac{1}{2}\mu _{\Lambda }^{2}+\omega R\right) \nabla _{\mu
 }\Lambda ^{\mu }+\left( \omega +\frac{\eta }{2}\right)
\left(\nabla _{\mu }R\right)\Lambda ^{\mu }+\eta \left(\nabla
_{\mu }\Lambda ^{\nu }\right)R_{\nu }^{\mu }=0.
\end{equation}

From Eq.~(\ref{3}) it follows that the four-potential of the
cosmological vector field satisfies the following wave equation:
\begin{equation}
\nabla _{\sigma }\nabla ^{\sigma }\Lambda ^{\mu }-\nabla ^{\mu
}\left( \nabla ^{\nu }\Lambda _{\nu }\right) -\left[ \left(
1+\frac{\eta }{4}\right) R_{\nu }^{\mu }+\frac{1}{4}\left( \omega
R+\frac{1}{2}\mu _{\Lambda }^{2}\right) \delta _{\nu }^{\mu
}\right] \Lambda ^{\nu }=0.
\end{equation}

Due to its antisymmetry, the massive vector field tensor
automatically satisfies the equations
\begin{equation}
\nabla _{\sigma }C_{\mu \nu }+\nabla _{\mu }C_{\nu \sigma }+\nabla
_{\nu }C_{\sigma \mu }=0.
\end{equation}

The matter energy momentum tensor $T_{\mu \nu }$ satisfies the
conservation law $\nabla^{\mu} T_{\mu\nu}=0$, which can be verified
by taking the covariant divergence of Eq.~(\ref{field}).

\section{Cosmological applications}

To investigate the cosmological implications of the massive vector-metric
theory we adopt the flat Robertson-Walker metric for a homogeneous and
isotropic Universe, given by
\begin{equation}
ds^{2}=dt^{2}-a^{2}(t)\left( dx^{2}+dy^{2}+dz^{2}\right) ,
\end{equation}
where $a(t)$ is the scale factor. The observed isotropy and homogeneity of
the Universe requires that the massive vector field is a function of the
cosmological time only. Hence we assume that the potential $\Lambda ^{\mu }$
has only one non-zero component, $\Lambda ^{\mu }=\left( \Lambda
^{0}(t),0,0,0\right) $. The function $\phi $ is given by $\phi (t)=\Lambda
_{0}(t)\Lambda ^{0}(t)$.

The non-zero components of the Ricci tensor and the
Ricci scalar are given by $R_{00}=-3\ddot{a}/a$, $R_{\alpha \alpha }=\left( a%
\ddot{a}+2\dot{a}^{2}\right) \delta _{\alpha \alpha }$, $\alpha =1,2,3$ and $R=-6\left( \ddot{%
a}/a+\dot{a}^{2}/a^{2}\right) $, respectively. Moreover, we have
$\nabla _{\alpha }\nabla _{\beta }\left(
\Lambda ^{\alpha }\Lambda ^{\beta }\right)=\ddot{\phi}+3\phi \ddot{a%
}/a+6\phi \dot{a}^{2}/a^{2}+6\dot{\phi}\dot{a}/a$, $\nabla
_{\sigma }\nabla ^{\sigma }\left( \Lambda
_{0}\Lambda _{0}\right)=\ddot{\phi}+3\dot{\phi}\dot{a}%
/a-6\phi \dot{a}^{2}/a^{2}$ and $\nabla _{\sigma }\nabla _{0
}\left( \Lambda _{0}\Lambda ^{\sigma
}\right)=\ddot{\phi}+3\dot{\phi}\dot{a}/a-3\phi \dot{a}%
^{2}/a^{2}$.

We assume that the matter content of the Universe consists of the
massive cosmological vector field $C_{\mu \nu }$ and ordinary
matter in form of pressureless dust with density $\rho $. The
conservation of the energy-momentum tensor then gives $\rho =\rho
_0/a^3$, where $\rho _0$ is a constant of integration.

Hence, the gravitational field equations and the equation of motion
of the vector field become
\begin{equation}
\left[ 1-\left( \omega -\eta \right) \phi \right] \frac{\dot{a}^{2}}{a^{2}}%
-\left( 2\omega +\eta \right) \phi \frac{\ddot{a}}{a}+\left( \omega +\frac{%
\eta }{2}\right) \dot{\phi}\frac{\dot{a}}{a}+\frac{\mu _{\Lambda }^{2}}{12}%
\phi =\frac{8\pi G_0}{3}\frac{\rho _0}{a^3},  \label{eq1}
\end{equation}
\begin{equation}
\left[ 2+\left( 2\omega +3\eta \right) \phi \right]
\frac{\ddot{a}}{a}+\left[ 1+\left( \omega +3\eta \right) \phi
\right] \frac{\dot{a}^{2}}{a^{2}}+\left(
\omega +\frac{\eta }{2}\right) \ddot{\phi}+3\left( \omega +\eta \right) \dot{%
\phi}\frac{\dot{a}}{a}-\frac{\mu _{\Lambda }^{2}}{4}\phi =0,
\end{equation}
\begin{equation}
\left( 2\omega +\eta \right) \frac{\ddot{a}}{a}+2\omega \frac{\dot{a}^{2}}{%
a^{2}}=\frac{\mu _{\Lambda }^{2}}{6}.  \label{eq2}
\end{equation}

Eliminating with the use of Eq.~(\ref{eq2}) the second derivative
of the scale factor from Eq.~(\ref{eq1}) gives
\begin{equation}
\left[ 1+\left( \omega +\eta \right) \phi \right] \frac{\dot{a}^{2}}{a^{2}}%
+\left( \omega +\frac{\eta }{2}\right)
\dot{\phi}\frac{\dot{a}}{a}-\frac{\mu _{\Lambda }^{2}}{12}\phi
=\frac{8\pi G_0}{3}\frac{\rho _0}{a^3}.  \label{eq3}
\end{equation}

The general solution of Eq.~(\ref{eq2}) can be represented in an
integral form as
\begin{equation}
t-t_{0}=\int \frac{da}{\sqrt{\frac{\mu _{\Lambda }^{2}}{6\left(
4\omega +\eta \right) }a^{2}+a_0^2a^{-\frac{4\omega }{2\omega
+\eta }}}},
\end{equation}
where $t_{0}$ and $a_0$ are arbitrary constants of integration.

In the limit of large $a$ and for $4\omega /(2\omega +\eta )>0$
the scale factor is given by
\begin{equation}
a=\exp \left[ H\left(t-t_0\right)\right],\qquad H={\rm constant},
\end{equation}
corresponding to a de Sitter type exponentially accelerating phase
for the expansion of the Universe. The constant $H$ is expressed
in terms of the mass of the cosmological vector field as
\begin{equation}\label{H}
H^{2}=\frac{\mu _{\Lambda }^{2}}{6\left( 4\omega +\eta \right) }=\frac{%
\Lambda }{3},
\end{equation}
where $\Lambda $ is the cosmological constant, which is generated
due to the presence of the massive cosmological vector particle,
and whose numerical value can be determined from observations.

For arbitrary times the exact form of the scale factor is given by
\begin{equation}\label{scale}
a(t)=a_{0}\sinh ^{n}\left[ \beta \left( t-t_{0}\right) \right],
\end{equation}
where we have denoted
\begin{equation}
n=\frac{2\omega +\eta }{4\omega +\eta },
\end{equation}
and
\begin{equation}
 \beta =\frac{\left( 4\omega +\eta \right) H}{\left( 2\omega
+\eta \right)}=\frac{H}{n},
\end{equation}
respectively.

With the use of Eq.~(\ref{scale}) it follows that the massive
cosmological vector field satisfies the evolution equation
\begin{multline}
\dot{\phi} = \beta \left[ \tanh \beta \left( t-t_{0}\right)
-\frac{2\left(
\omega +\eta \right) }{4\omega +\eta }\coth \beta \left( t-t_{0}\right) %
\right] \phi -\frac{2H}{2\omega +\eta }\coth \beta \left(
t-t_{0}\right)\\
+\frac{16\pi G_{0}\rho _{0}}{3\left( 2\omega +\eta \right) a_{0}^{3}H}%
\frac{\sinh ^{1-3n}\beta \left( t-t_{0}\right) }{\cosh \beta
\left( t-t_{0}\right) },
\end{multline}
with the general solution given by
\begin{multline}
\phi (t)=\frac{\cosh \left[ \beta \left( t-t_{0}\right) \right] }{\sinh ^{%
\frac{2\left( \omega +\eta \right) }{4\omega +\eta }}\left[ \beta
\left( t-t_{0}\right) \right] }\biggl\{ B
+\frac{16\pi G_{0}\rho_{0}}
{3\left( 4\omega +\eta \right) a_{0}^{3}H^{2}}\tanh \left[
\beta \left( t-t_{0}\right) \right]\\ -\frac{2}{4\omega +\eta
}F\left[ \beta \left( t-t_{0}\right) \right] \biggr\} ,
\end{multline}
where $B$ is an arbitrary constant of integration, and
\begin{equation}
F(x)=\int \frac{dx}{\sinh ^{m}\left( x\right)},
\end{equation}
where
\begin{equation}
m=\frac{ 2\omega -\eta  }{ 4\omega +\eta }.
\end{equation}

During the pure de Sitter phase, with the effect of the ordinary
matter neglected, the equation describing the dynamics of the time
varying cosmological vector field $\phi $ is
\begin{equation}
\dot{\phi}=\frac{2\omega -\eta }{2\omega +\eta }H\phi
-\frac{2}{2\omega +\eta }H,
\end{equation}
with the general solution
\begin{equation}
\phi \left( t\right) =\phi _{0}\exp \left( \frac{2\omega -\eta
}{2\omega +\eta }Ht\right) +\frac{2}{2\omega -\eta }\left[ 1-\exp
\left( \frac{2\omega -\eta }{2\omega +\eta }Ht\right) \right] ,
\end{equation}
where $\phi _{0}$ is the initial value of the field at the initial time $%
t=t_{0}=0$, $\phi \left( 0\right) =\phi _{0}$.

In the case of a constant field $\phi =\phi _0=$constant, $H$ and
$\phi $ are related by
\begin{equation}
H^{2}=\frac{\mu _{\Lambda }^{2}}{12}\frac{\phi _0}{1+\left( \omega
+\eta \right) \phi _0},
\end{equation}
from which we find the constant massive cosmological vector field
as
\begin{equation}
\phi _0=\frac{4\Lambda /\mu _{\Lambda }^2}{1-4\left(\omega +\eta
\right)\Lambda /\mu _{\Lambda }^2}=\frac{2}{2\omega -\eta }.
\label{phizero}
\end{equation}

\section{PPN constraints on the model parameters}

The numerical values of the coupling coefficients $\omega $ and
$\eta $ and the initial value $\phi _0$ of the cosmological vector
field can be constrained by using solar system observations.
Vector tensor models generate observable effects in light
deflection and retardation experiments, planetary perihelion
advance, orbiting gyroscope precession, non-secular terms in
planetary and satellite orbits, geophysical phenomena etc. These
effects can be described in terms of the dimensionless parameters
$\alpha $, $\beta $ and $\gamma $, which parameterize deviations
with respect to standard general relativity. Actually, the
parameter $\alpha $ can be settled to the unity due to the mass
definition of the system itself~\cite{will}.

The quantity $\gamma -1$ measures the degree to which gravity is
not a purely geometric effect, and it is affected by other fields.
Measurements of the frequency shift of the radio photons to and
from the Cassini spacecraft as they passed near the Sun give the
result $\gamma =1+\left( 2.1\pm 2.3\right) \times
10^{-5}$~\cite{ber03}. The value of the parameter $4\beta -\gamma
-3$ can be constrained from Lunar Laser Ranging, with the
observational result $4\beta -\gamma -3=-\left( 0.7\pm 1\right)
\times 10^{-3}$~\cite{will96}. For the cosmological time variation
of the effective gravitational constant we adopt the value
$\dot{G}/G=10^{-14}$ yr$^{-1}$~\cite{uzan}. As for the present
value of the Hubble constant we take $H=70\,{\rm km/s/Mpc}$.

Therefore, the three free parameters of our model $\left(
\phi_{0},\omega ,\eta \right)$ can be obtained from the following
non-linear system of algebraic equations:
\begin{equation}
\gamma -1=\frac{2\omega \left( 1+\omega -\eta /2\right)
\bar{\phi}}{1-\omega \left( 1+4\omega \right)
\bar{\phi}}=2.1\times 10^{-5},
\label{gamma}
\end{equation}
\begin{multline}
4\beta -\gamma -3=\frac{\left( \gamma -1\right) \left( 1-\omega \bar{\phi}%
\right) }{\omega \bar{\phi}}\left\{ 1+\frac{\gamma \left( \gamma -2\right) }{%
\frac{1}{2}\left( \gamma +1\right) -\frac{\eta \left( \gamma -1\right) }{%
4\omega }+\left[ \frac{3\omega \left( \gamma -1\right)
}{2}+\frac{\eta \left( \gamma -3\right) }{4}\right]
\bar{\phi}}\right\}\\ =-0.7\times 10^{-3},
\label{beta}
\end{multline}
\begin{equation}
\frac{1}{H}\frac{\dot{G}}{G}=\frac{\frac{3\omega \left( \gamma -1\right) }{2}%
+\frac{\eta \left( \gamma -3\right) }{4}}{2\omega +\eta
}\frac{\left[ \left(
2\omega -\eta \right) \phi _{0}-2\right] \exp \left( \frac{2\omega -\eta }{%
2\omega +\eta }\right) }{\left\{ \frac{1}{2}\left( \gamma +1\right) -\frac{%
\eta \left( \gamma -1\right) }{4\omega }+\left[ \frac{3\omega
\left( \gamma
-1\right) }{2}+\frac{\eta \left( \gamma -3\right) }{4}\right] \bar{\phi}%
\right\} }=1.40\times 10^{-4},
\label{newtong}
\end{equation}
where we denoted by $\bar{\phi}$ the present day value of the
cosmological vector field,
\begin{equation}
\bar{\phi}=\phi \left( \frac{1}{H}\right) =\left( \phi
_{0}-\frac{2}{2\omega
-\eta }\right) \exp \left( \frac{2\omega -\eta }{2\omega +\eta }\right) +%
\frac{2}{2\omega -\eta }.
\end{equation}

These three equations are highly non-linear, and approximate
solutions can only be obtained by the means of numerical methods.

\begin{figure}[h]
\centering
\includegraphics[width=.48\textwidth]{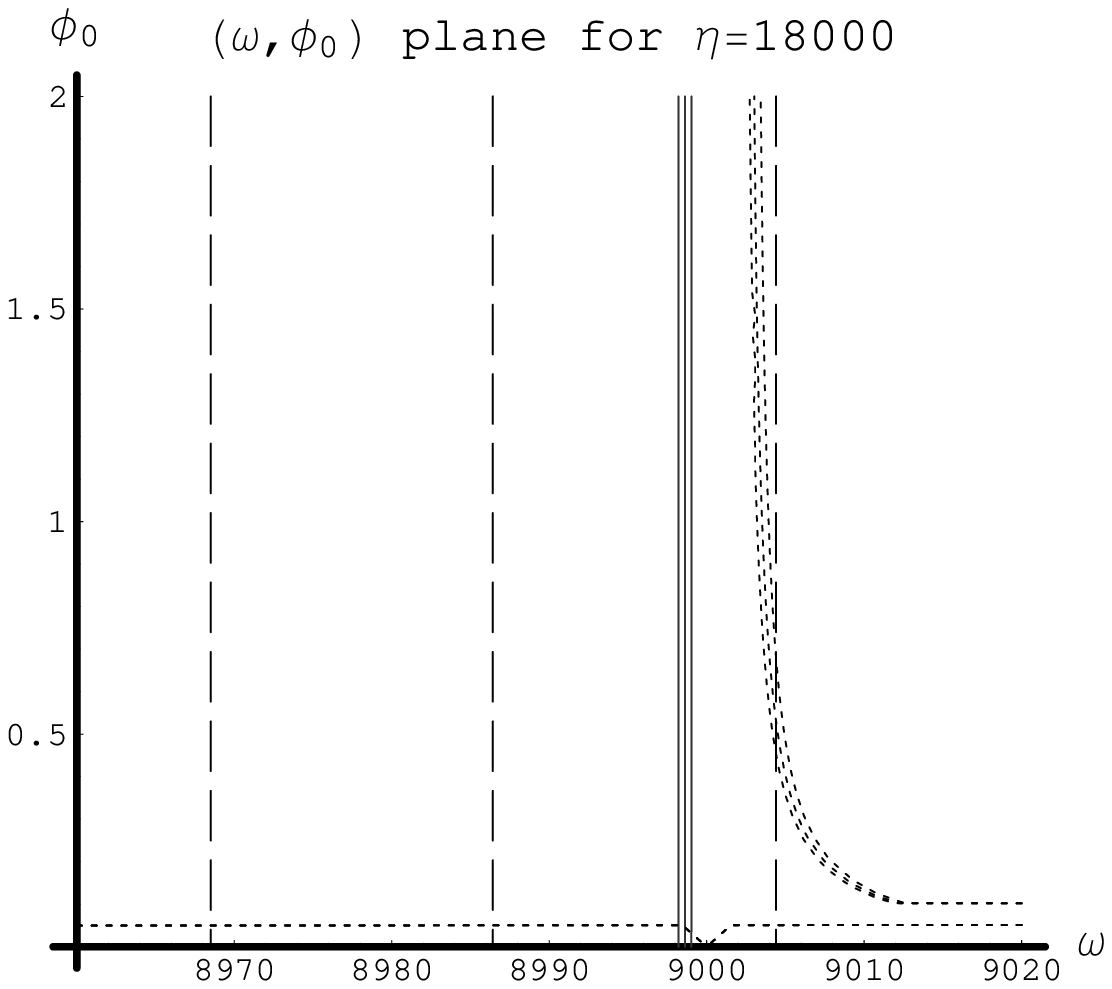}
\includegraphics[width=.48\textwidth]{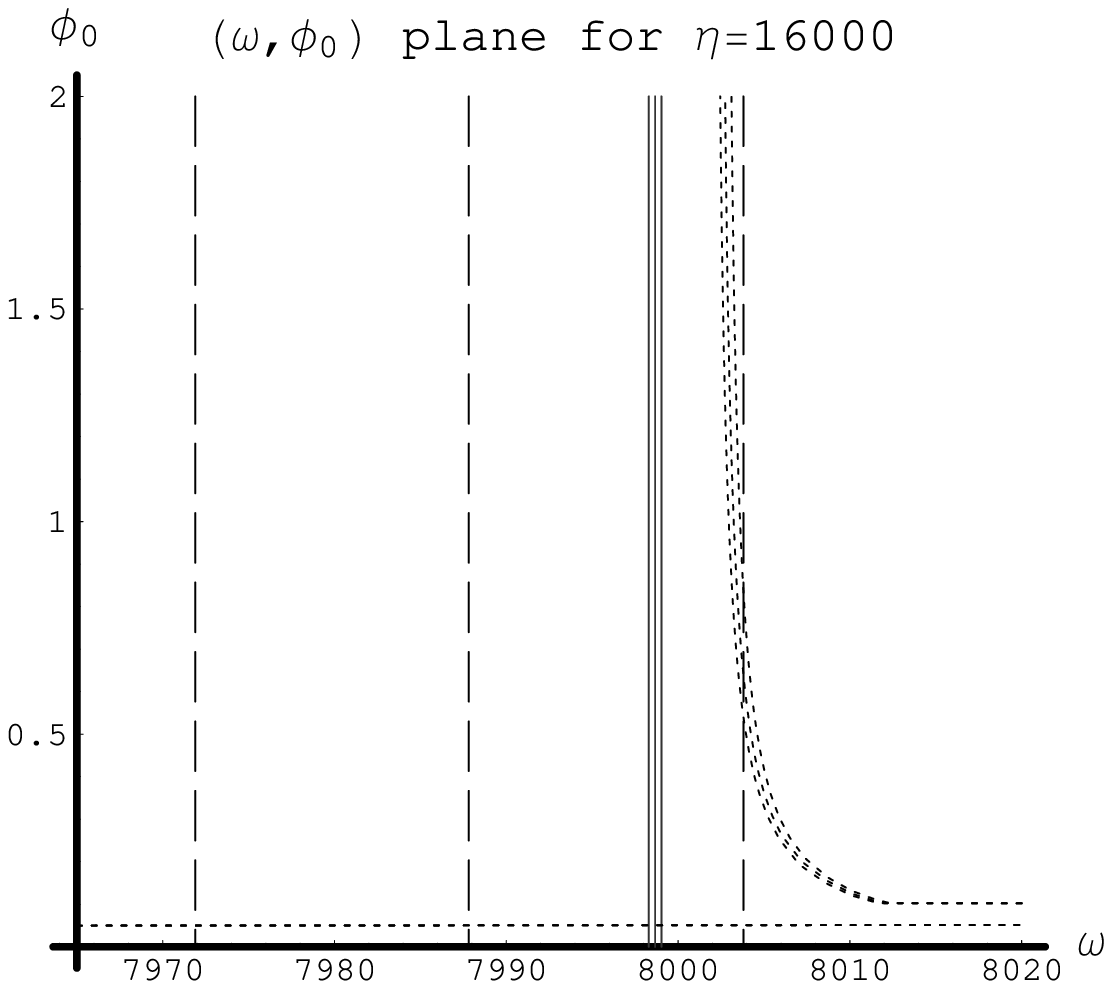}
\caption{These figures represent the $(\omega,\phi_0)$-planes for $\eta=18000$ and
         $\eta=16000$. The three straight lines in the middle represent the allowed
         range of the parameter $\gamma-1 = (2.1 \pm 2.3)\times 10^{-5}$, as given
         by Eq.~(\ref{gamma}). The dashed lines represent the allowed range of
         $4\beta-\gamma-3$ via Eq.~(\ref{beta}). The right dashed lines correspond
         to the value $4\beta-\gamma-3=-(0.7-1)\times 10^{-3} = +0.3\times 10^{-3}$
         and implies that the model strongly favors positive values. Consequently,
         a further decrease on the error bars of $4\beta-\gamma-3$ can either rule
         out or further strengthen the present model. Lastly, the dotted lines describe
         the allowed parameter range for the `cosmological' equation~(\ref{newtong}),
         where we used $H=70\pm 10\,{\rm km/s/Mpc}$.
         It is evident that there is no parameter range where all three equations are
         satisfied to given accuracy, in fact one has to search for parameters that
         minimize the error.}
\label{n18000}
\end{figure}

\begin{figure}[htp]
\centering
\includegraphics[width=.48\textwidth]{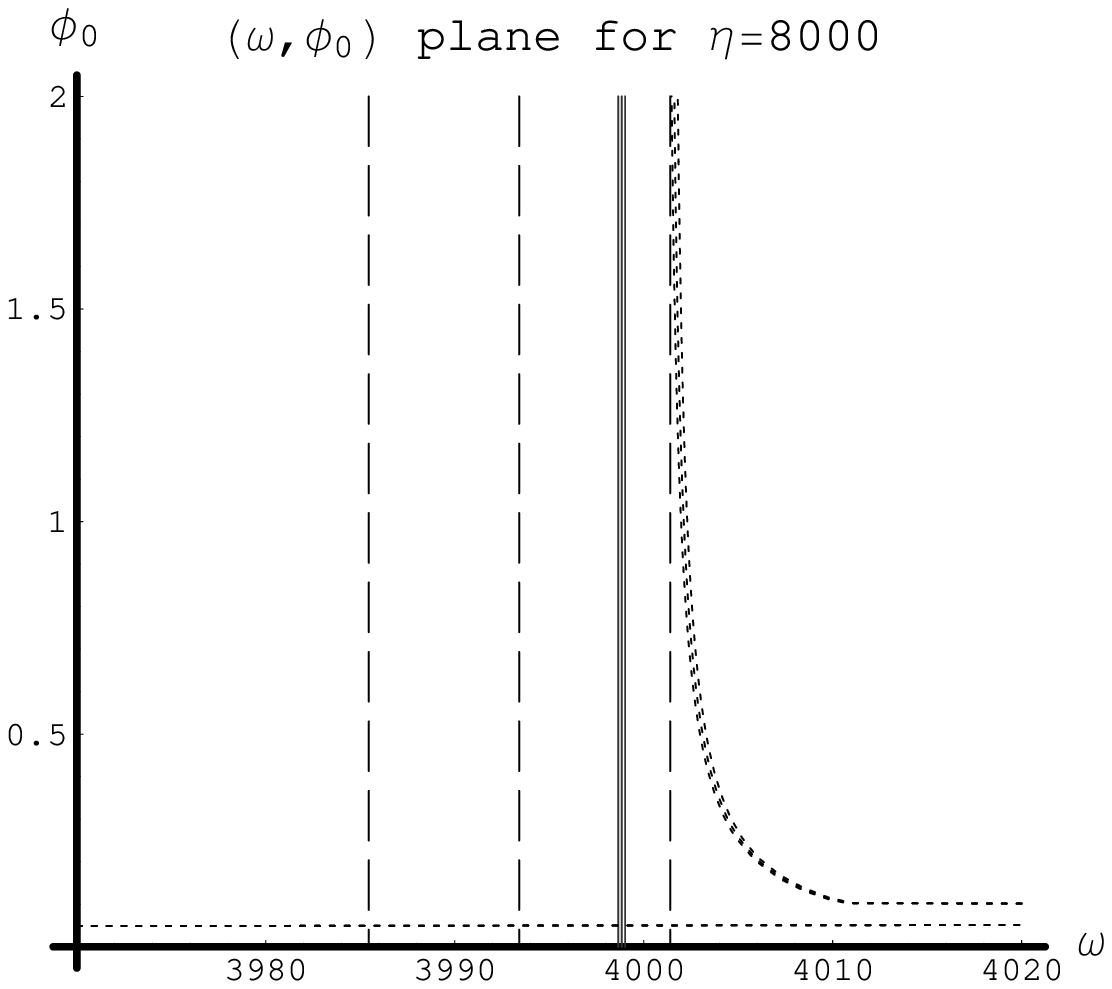}
\includegraphics[width=.48\textwidth]{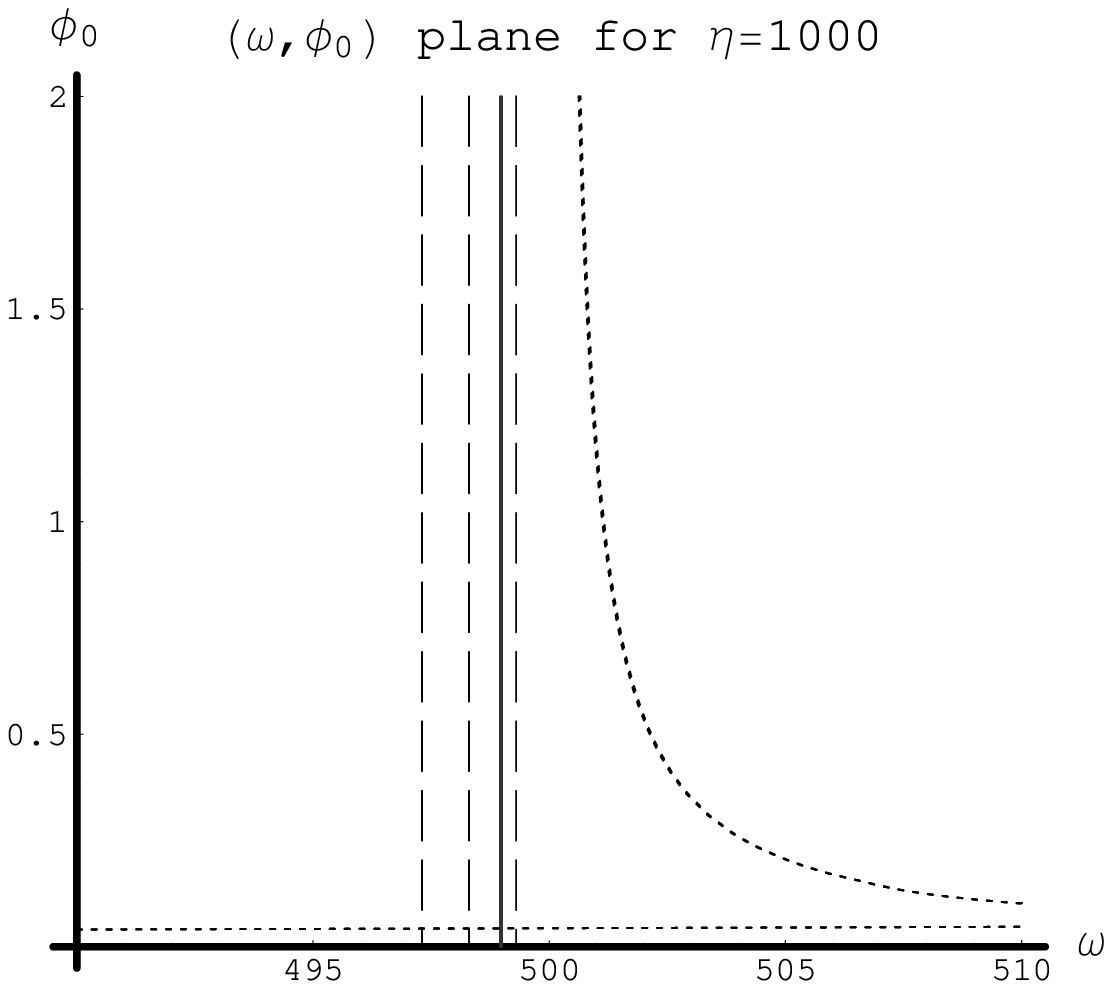}
\caption{Using the same coding as in the previous figure, it should be noted that
         the possibility of fitting the three parameters becomes more difficult.
         Already for $\eta=8000$ the `cosmological' range of parameters (dotted) lies outside
         the allowed range of parameter as implied by Eq.~(\ref{beta}) (dashed). Hence, it
         is even more difficult to fit the three parameters simultaneously. A further
         increase of $\eta$ larger than $18000$ has the same effect, the situation
         does not improve.}
\label{n8000}
\end{figure}

To constrain the numerical values for the three free parameter, we
firstly consider the $(\omega,\phi_0)$-plane for given values of
the parameter $\eta$ in Figures~\ref{n18000} and~\ref{n8000}.

\begin{figure}[hbp]
\centering
\includegraphics[width=.48\textwidth]{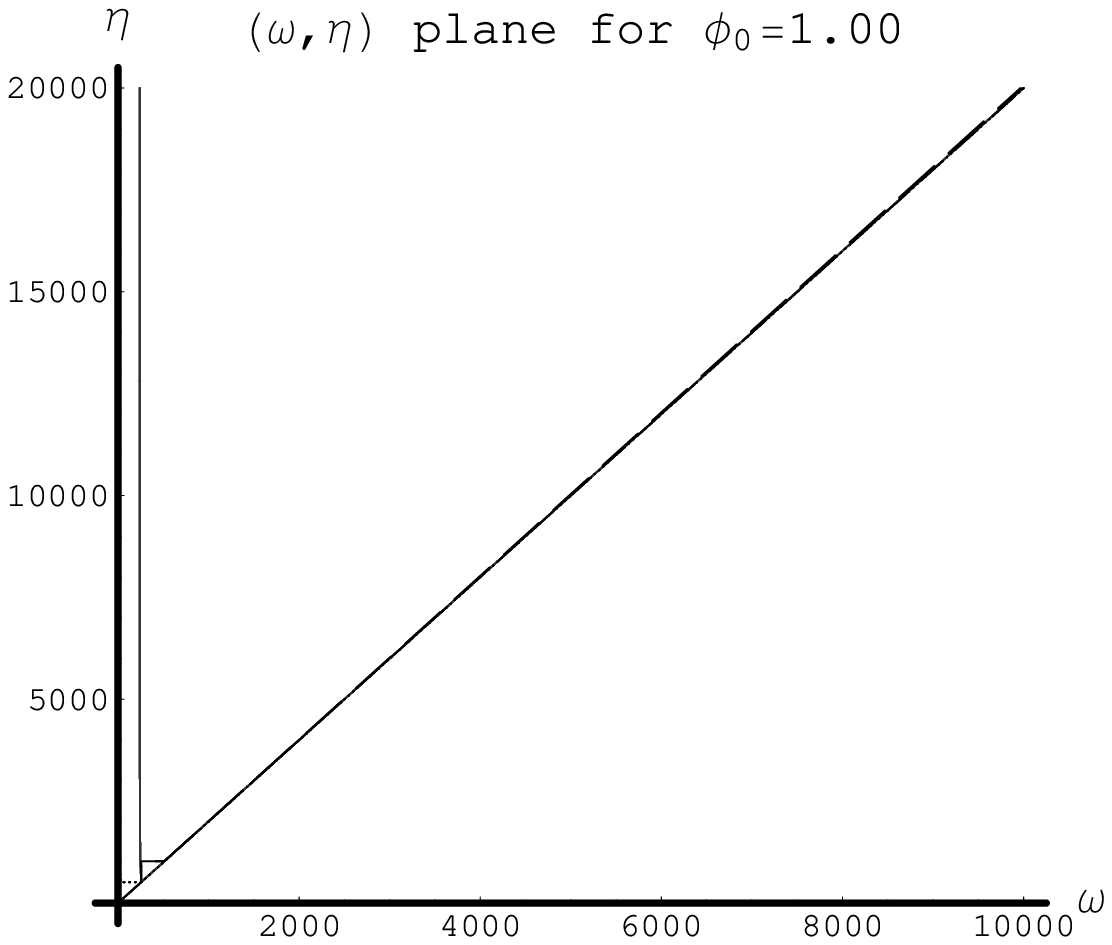}
\includegraphics[width=.48\textwidth]{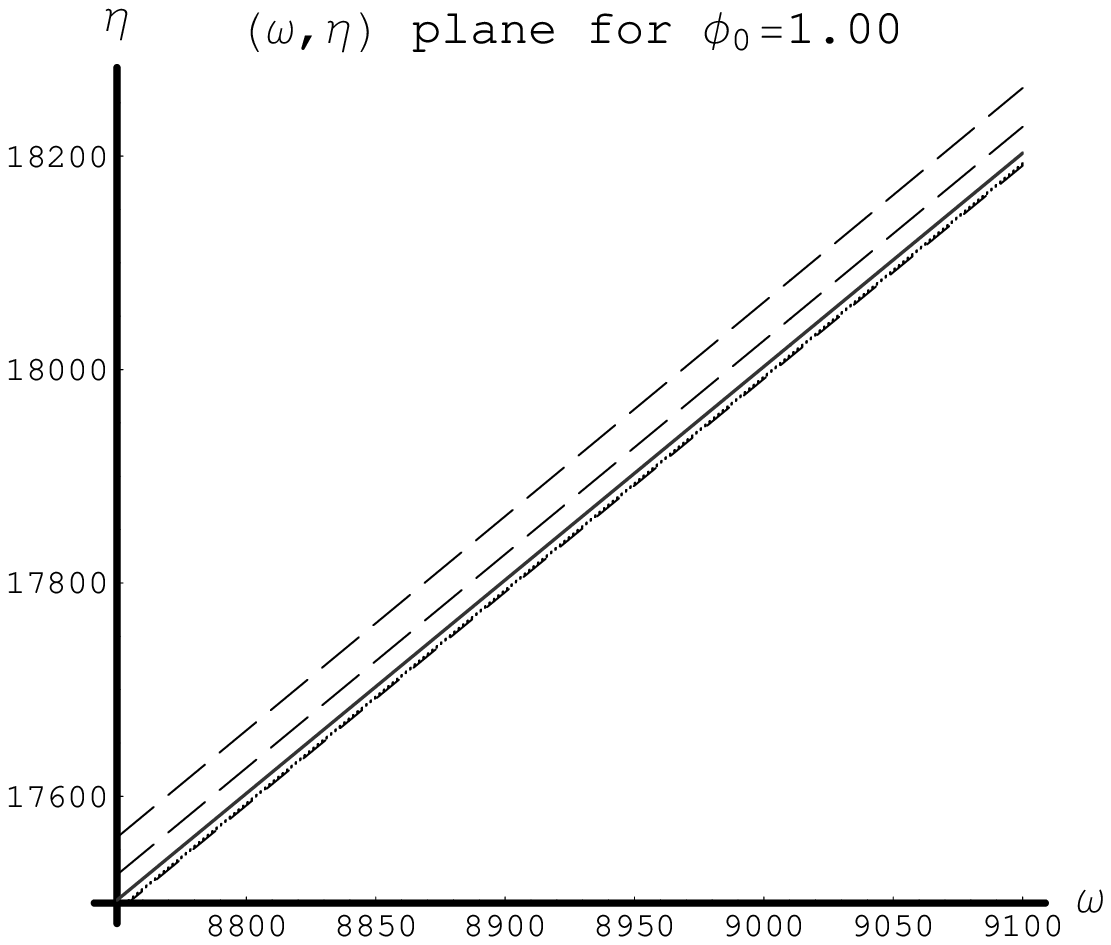}
\caption{As in the above figures, straight lines show the $\gamma-1$ range, dashed lines
         the $4\beta-\gamma-3$ range, and dotted lines the cosmological constraint. On the
         left-hand side, where $0<\omega<10000$, the different lines cannot be distinguished.
         The right-hand side ($8750<\omega<9100$) again indicates that one can find some
         approximate solution for the three parameters. An increase of the parameter $\phi_0$
         does not change the qualitative picture.}
\label{phi}
\end{figure}

The $(\eta,\phi_0)$-planes share features very similar to those of
the $(\omega,\phi_0)$-planes, therefore it is not particularly
insightful to also consider these. The reason behind this is that
the numerical fitting of the three parameters works best if
$2\omega-\eta$ is of the order $O(10^{-5})$. The smallness of this
difference has its roots in the Newtonian limit of the theory,
which in case of a massless vector field reduces to
$2\omega=\eta$. Hence, this small deviation is due to the massive
vector field. However, let us consider the $(\omega,\eta)$-planes
for given $\phi_0$ in Figure~\ref{phi}.

For example, one possible set of numerical values for the free parameters
of the model is given by $\phi_{0}=1.097999982$, $\omega =9000.000069565$
and $\eta=18000.0001381$, respectively. These values satisfy the
general constraints on the coupling constants $2\omega -\eta >0$
and $4\omega +\eta >0$.

The first condition is implied by the positivity of $\phi_0$, see
Eq.~(\ref{phizero}), since $\phi$ is the square of the massive
vector field. The second condition follows from Eq.~(\ref{H}),
i.e.~positivity of the vector field mass and positivity of the
Hubble constant. The above values represent an approximate
solution of the system of constraint equations of the order of
$O\left(10^{-5}\right) $, $O\left( 10^{-4}\right) $ and
$O\left(10^{-4}\right) $. It is interesting to note that the
system of equation is quite sensitive towards a change of the
`cosmological' equation~(\ref{newtong}).

If, for the moment, we assume that the gravitational constant $G$
changes slower than given by the upper bound $\dot{G}/G =
10^{-14}\; {\rm yr}^{-1}$, for example an order of magnitude
slower, then also the fit of the `cosmological' equation improves
by an order of magnitude. Hence, we can conclude that the analyzed
theoretical model makes mild predictions regarding observation:

The gravitational constant should vary slower over time and the
parameter $4\beta-\gamma-3$, constrained by Lunar Laser Ranging,
should lie in the positive region. On the other hand, it should be
emphasized that our model would not be in good agreement with
$\dot{G}=0$, a constant effective gravitational ``constant''.

\section{ Discussions and final remarks}

From Eq.~(\ref{H}) we obtain for the mass of the cosmological
vector particle the expression
\begin{equation}
      \mu_{\Lambda} = \sqrt{6\left(4\omega +\eta\right)}\, H
      \approx 1.67 \times \sqrt{6\left(4\omega +\eta\right)}
      \times 10^{-63}\; {\rm g}.
\end{equation}
The upper limit for the mass of the ordinary photon obtained by
using a rotating torsion balance method is
$1.2\times 10^{-51}\; {\rm g}$~\cite{Luo03}. It should be noted
that the proposed massive vector field interacts only with gravity
and has no standard matter interactions. The existence of
a minimal mass in nature in the presence of a cosmological constant
has been discussed for example in~\cite{BoHa05}.

The mass of the cosmological massive vector particle may have been
generated during inflation. Inflation allows the emergence of
fields coherent over large distances from quantum fluctuations and
prevents dissipative effects, due to the absence of charged
particles. The conformal invariance of the U(1) gauge theory for
electromagnetism prevents the gravitational field from producing
photons. However, in the present model, due to the breaking of the
conformal invariance, via the coupling to gravity, the vector
particle interacting with gravity can acquire an effective mass.
Therefore, the effective mass for the massive vector particle
composing the cosmological gas filling the Universe, and
representing its main matter component, may have originated during
inflation~\cite{Tu88}. The mass of the cosmological vector
particle can also be acquired from the spontaneous breaking of
Lorentz symmetry in the context of field theories arising from
string field theory, or due to the introduction of a fundamental
minimal length in a trans-Planckian physics scenario~\cite{Be99}.

The vector field dominated Universe enters in a pure exponential
de Sitter phase when the condition $t>t_{acc}=1/\beta
=\left(2\omega +\eta \right)t_H/\left(4\omega +\eta \right)$ is
satisfied. With the use of the values of the parameters $\omega $
and $\eta $ obtained from the PPN solar system constraints we
obtain $t_{acc}\approx 2 t_H/3$, corresponding to a redshift of
$z_{acc}\approx 0$. In fact, the supernova data published by the 
High-$z$ Supernova Search Team and the Supernova Cosmology Project 
show that the transition from the decelerating to the accelerating 
phase occurred at redshifts smaller than $z=0.4$~\cite{TuRi02}. 
Therefore, the model predicts that the accelerated expansion phase 
of the Universe started only recently.

In conclusion, we have shown that the dark energy can be modeled
as a massive cosmological vector field filling the Universe. This
field may have originated during the inflationary period, when the
vector field may have acquired its mass, and it drives the late-time
acceleration of the Universe. All the parameters of the model can
be constrained from observational data.

\section*{Acknowledgments}

We would like to thank to anonymous referee for helpful
comments and suggestions, which helped us to greatly improve the
manuscript. The work of C.~G.~B.~was supported by research grant
BO 2530/1-1 of the German Research Foundation (DFG). The work of
T.~H.~was supported by the RGC grant No.~7027/06P of the government 
of the Hong Kong SAR.

\end{document}